\documentclass{article}
\usepackage{spconf}
\usepackage{cite}
\usepackage{amsmath,amssymb,amsfonts}
\usepackage{graphicx}
\usepackage{textcomp}
\usepackage{xcolor}

\usepackage{times}
\usepackage{soul}
\usepackage{url}
\usepackage[hidelinks]{hyperref}
\usepackage[utf8]{inputenc}
\usepackage[small]{caption}
\usepackage{graphicx}
\usepackage{amsthm}
\usepackage{mathrsfs} 
\usepackage{booktabs}
\usepackage{multirow}
\usepackage{algorithm}  
\usepackage{algorithmicx}  
\usepackage{algpseudocode}
\renewcommand{\algorithmicrequire}{\textbf{Input:}}  

\usepackage{tikz}
\usepackage{subcaption}
\usepackage{balance}

\usetikzlibrary{decorations.pathreplacing,calc}

\usepackage{listings}

\newtheorem{theorem}{Theorem}
\newtheorem{assumption}{Assumption}

\newtheorem{definition}{Definition}
\usepackage{environ}
\NewEnviron{iproof}[1][Proof]{\begin{IEEEproof}[\indent #1]\BODY\end{IEEEproof}}{}
\begin{document}

\title{S2 Reducer: High-Performance Sparse Communication to Accelerate Distributed Deep Learning}

\name{Keshi Ge, Yongquan Fu, Yiming Zhang, Zhiquan Lai, Xiaoge Deng, Dongsheng Li }
\address{College of Computer, National University of Defense Technology,\\
\textit{{\{gekeshi, yongquanf, zhangyiming, zqlai, dengxg, dsli\}@nudt.edu.cn}}}


\maketitle

\begin{abstract}
Distributed stochastic gradient descent (SGD) approach has been widely used in large-scale deep learning, and the gradient collective method is vital to ensure the training scalability of the distributed deep learning system. Collective communication such as AllReduce has been widely adopted for the distributed SGD process to reduce the communication time. However, AllReduce incurs large bandwidth resources while most gradients are sparse in many cases since many gradient values are zeros and should be efficiently compressed for bandwidth saving. To reduce the sparse gradient communication overhead, we propose Sparse-Sketch Reducer (S2 Reducer), 
a novel sketch-based sparse gradient aggregation method with convergence guarantees. 
S2 Reducer reduces the communication cost by only compressing the non-zero gradients with count-sketch and bitmap, and enables the efficient AllReduce operators for parallel SGD training.
We perform extensive evaluation against four state-of-the-art methods over five training models. Our results show that S2 Reducer converges to the same accuracy, reduces 81\% sparse communication overhead, and achieves 1.8$ \times $ distributed training speedup compared to state-of-the-art approaches.


\end{abstract}


\section{Introduction}\label{sec:intro}
Deep neural networks (DNN) have been widely used in computer vision~\cite{he2016deep,huang2017densely,dosovitskiy2020image}, speech recognition~\cite{gulati2020conformer,xu2021self}, and natural language processing (NLP) tasks~\cite{dai2019transformer,mehta2020delight}, A recent trend is that the state of the art (SOTA) results in these areas are achieved by large DNN models. Training large models needs long training periods and high computing costs. Distributed deep learning process uses many computing devices to reduce the training time~\cite{ben2019demystifying}, which parallelizes the computation of the model. Data-parallel Stochastic Gradient Descent (SGD)~\cite{li2014scaling, fan2021dapple} is the most popular approach to implement the distributed deep learning paradigm, because of its flexibility and well support by the deep learning toolkit, such as TensorFlow~\cite{abadi2016tensorflow}, PyTorch~\cite{paszke2019pytorch} and DeepSpeed~\cite{rasley2020deepspeed}. The data-parallel SGD iterates over two steps as follows: (1) Each device conducts the forward and backward computation of the DNN model in parallel with sampled mini-batch data, and produces the local gradient; (2) Each device collects all the gradients from other devices and calculates the average of these gradients to update the model parameters.

In distributed SGD, iterative gradient aggregation is conducted among devices, which costs heavy network bandwidth resources. The bandwidth costs increase fast with larger models and even delay the training process~\cite{alistarh2017qsgd, lin2018deep}. To reduce the communication time, AllReduce, an efficient and widely-supported collective communication operation~\cite{jeaugey2017nccl}, aggregates the gradients on GPU directly~\cite{peng2019generic, cho2019blueconnect}. However, these implementations ignore the fact that AllReduce is not suitable for sparse gradient, in which most of the elements are zero. Sparse gradient are quite prevalent, \textit{e.g.}, the gradient sparsity of embedding layers in some NLP models exceed 94\%~\cite{fei2021efficient}. Some communication-efficient SGD, such as block-wise gradient sparsification~\cite{vogels2019powersgd, fei2021efficient} also yield sparse gradient. Thus, a large volume of zeros is transmitted~\cite{kim2019parallax}. What's worse, the compressed sparse gradient, which typically contains index and value of non-zeros of sparse gradient, cannot be added up directly, unlike what the dense gradients do in AllReduce, because the indices of non-zero values among GPUs are maybe different. The current implementation of sparse collective communication in deep learning frameworks, such as PyTorch, requires two All-Gather operations to synchronize the indices and values of sparse gradient separately. However, each device has to collect all the other sparse gradients before summing up, which has poor communication efficiency. Some other sparse-oriented systems, such as SparCML~\cite{renggli2019sparcml} and Parallax~\cite{kim2019parallax} introduce dynamic sparse collective communication methods to spend less transmission time, yet they have to use the inefficient AllGather or convert the sparse input to dense. Although OmniReducer~\cite{fei2021efficient} devises a communication library that also identifies and aggregates the non-zero blocks, it still wastes bandwidth to transfer zero blocks. While recent work~\cite{ivkin2019communication} also enables AllReduce on a compressed gradient using count-sketch, it can only approximately find the Top-K items from the dense gradient, which does not return estimate values from the sketch directly.

In this paper, we propose Sparse-Sketch Reducer(S2 Reducer), an aggregation method to collect sparse gradients efficiently. 
The key idea of our approach is compressing the non-zeros with a novel sparse-sketch, which supports AllReduce operation and guarantees the convergence of data-parallel SGD.
The sparse-sketch compresses the non-zeros with a count-sketch and a bitmap, which can be synchronized with AllReduce. S2 Reducer produces an unbiased estimation when querying the gradient from the aggregated sketch, which guarantees the convergence theoretically and experimentally.


We have realized S2 Reducer\footnote{\url{https://github.com/GeKeShi/Sparse_Sketch_Reducer}}
and conducted extensive experiments on various CNN and NLP models in a cluster with up to 16 GPUs. 
The results demonstrate that 
our method achieves the same level of model accuracy against the baseline approaches
while achieving up to  1.8$\times$ speedup compared with the SOTA method and reducing 81\% sparse communication time.

\section{Background and Motivation}

\begin{figure}[t]
	\renewcommand{\arraystretch}{1.0}
	\centering
	\begin{subfigure}[b]{0.45\linewidth}
		\includegraphics[width=\linewidth]{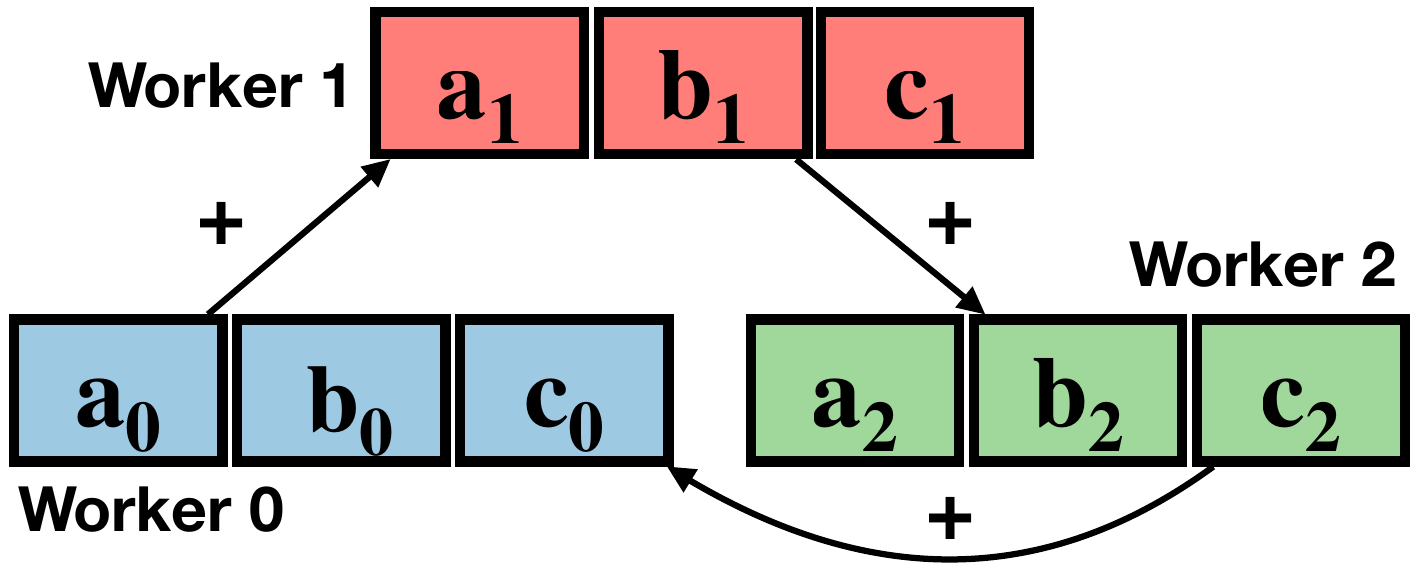}
		\caption{AllReduce w.r.t dense gradient.}
	\end{subfigure}
	\begin{subfigure}[b]{0.45\linewidth}
		\includegraphics[width=\linewidth]{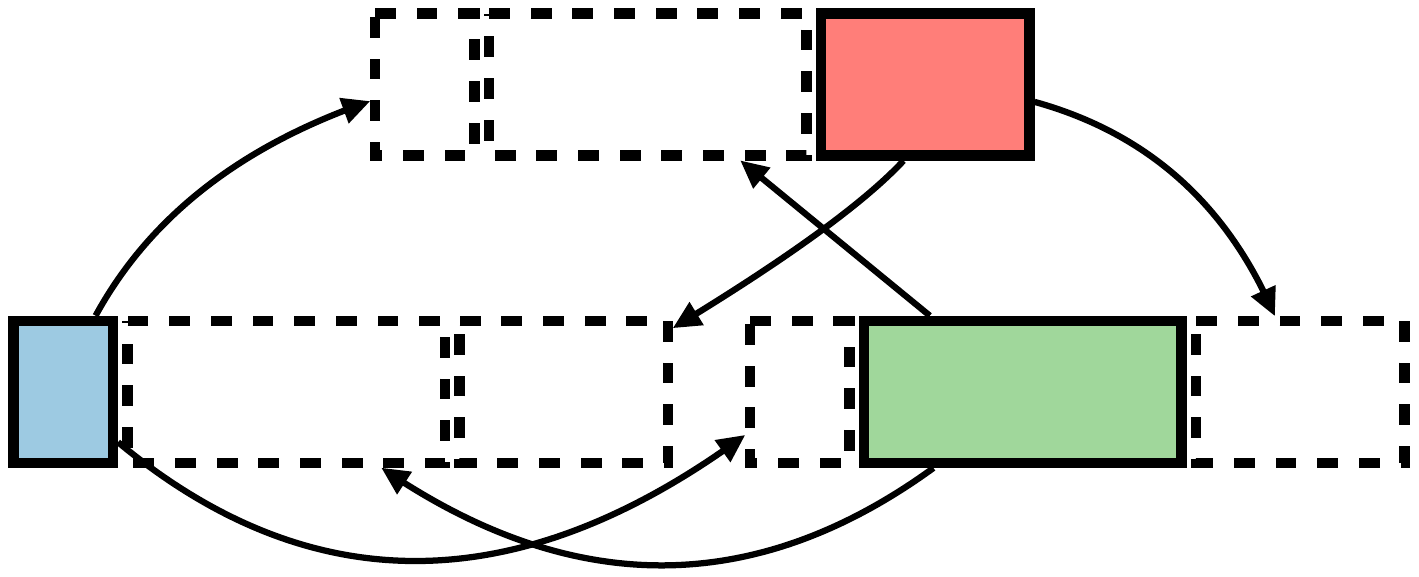}
		\caption{AllGather w.r.t sparse gradient.}
	\end{subfigure}
	\caption{Gradient transmission in AllReduce and AllGather. (a) A step of ring AllReduce. (b) Sparse gradients from other workers are gathered to concatenate.}
	\label{fig:collective_comm}
\end{figure}

\subsection{AllReduce Opertaion}

In distributed SGD, worker node $i\in\{1,\dots, W\}$ samples a mini-batch data, computes local gradient $g $ with back-propagation. Typically, they synchronize these gradients with \textit{ring AllReduce} to update the model parameters. Most high-performance distributed DNN training systems deploy AllReduce to achieve the speed records~\cite{ben2019demystifying, peng2019generic, fan2021dapple}.
Fig.~\ref{fig:collective_comm} (a) illustrates an example of a reduce step in the ring AllReduce for three devices. Each GPU worker partitions its gradient into $ W $ parts, such as $ \{a_0, b_0, c_0\} $, then each part are reduced across all workers by $ W-1 $ summation steps. Thus, only the data structure that can be added up supports AllReduce. 

\subsection{Saprse Gradient \& AllGather} 
Given a gradient vector of size $ d $, block-wise sparsification transform it to the matrix $ g^{b \times \frac{d}{b}} $, where $ \frac{d}{b} $ is the block size. Block-wise Top-K computes the $ \ell_2 $ norm of each block and finds the largest $ K $ blocks. The elements in these blocks are selected to communicate. 

The sparse gradients from the embedding layers are also block-wise. Each embedding vector encodes a word. In one SGD iteration, only a subset of word embedding vectors are selected as mini-batch and produce corresponding gradient vectors, each vector is a block.

These sparse gradients are typically compressed with the indices and values of the non-zeros. However, The difference among the indices in different workers prevents the aggregation method from using the efficient AllReduce. A compromising approach is AllGather. As shown in Fig.~\ref{fig:collective_comm} (b), workers collect non-zeros from others to synchronize. However, it is much slower than AllReduce, because it has to send extra traffic~\cite{fei2021efficient}. 
\subsection{Sketch Compression}\label{subsec2.3}


Count-sketch is one of the most popular communication traffic statistical algorithms in distributed systems, thanks to its mergeability property~\cite{agarwal2013mergeable}. 

The block-wise sparsity of gradient provides an opportunity to compress the sparse data with such a sketch structure that can be merged in AllReduce. The small volume of non-zero values can be stored with a fixed-sized count-sketch. Elements in the same block share the same row index. Thus, the indices of the sparse gradient can be encoded with a mergeable bitmap. Then, the irregular sparse gradients, like these in Fig.~\ref{fig:collective_comm} (b), are transformed to a data structure that can be added to another in a way that is shown in Fig.~\ref{fig:collective_comm} (a).
\section{Sketch Based Sparse Communication}
\begin{figure}[t]
	\centering
	\includegraphics[width=0.8\linewidth]{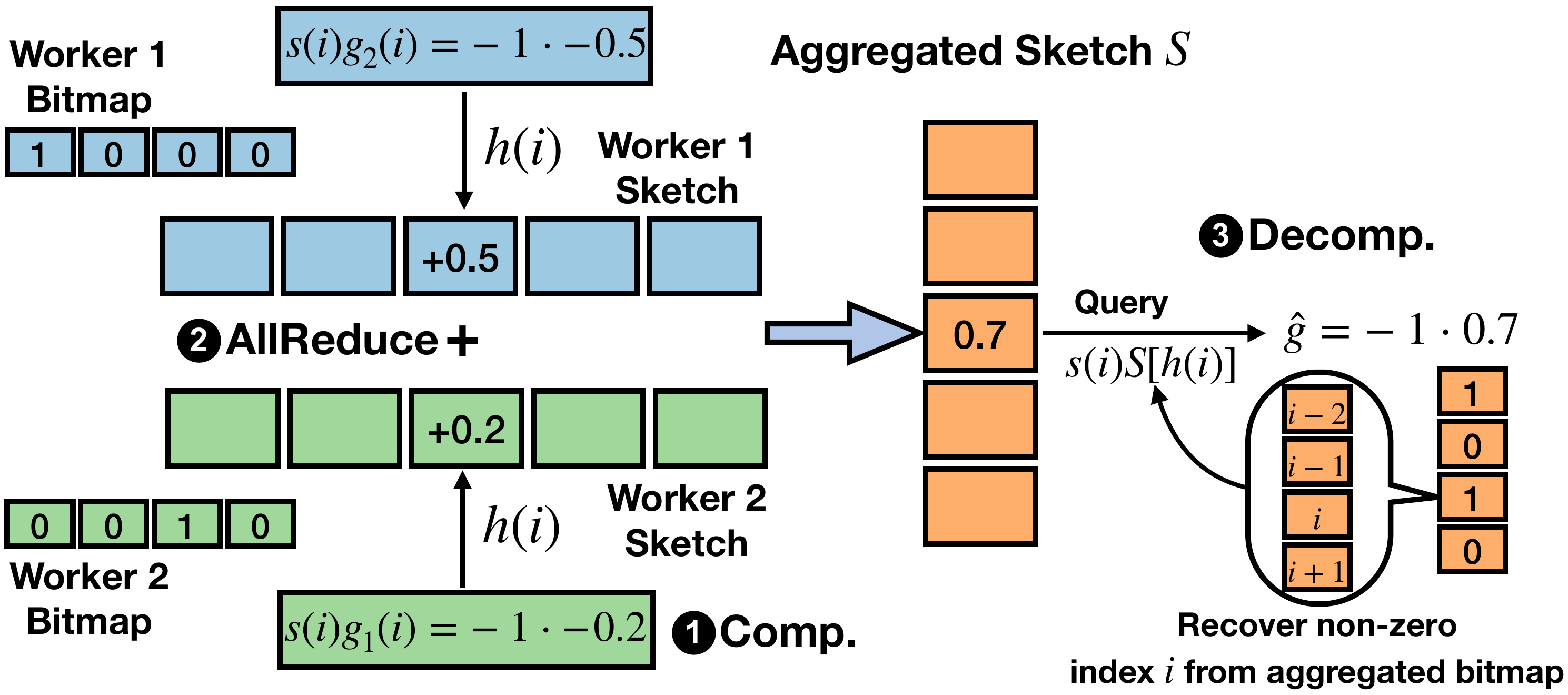}
	\caption{An example of the workflow of S2 Reducer.}
	\label{fig:sparsesketch}
\end{figure}
In this section, we devise Sparse-Sketch Reducer. The proposed sparse-sketch addresses two challenges in sparse gradient communication: (1) Enabling All-Reduce operation while compressing the non-zeros; (2) Guaranteeing the convergence of data-parallel SGD.

\subsection{Sparse-Sketch Reducer}{\label{subsec:sparse_sketch}}

\vspace{1mm}\noindent\textbf{Communication Workflow.}
The novel sparse-sketch only compresses the non-zeros and can be collected with efficient AllReduce.
We illustrate the workflow of S2 Reducer as Fig.~\ref{fig:sparsesketch}. Firstly, the indices are encoded with a bitmap, and non-zero values are inserted into a count-sketch. Then, both the bitmap and count-sketch are aggregated by AllReduce. Finally, the aggregated sparse gradient is decompressed to update the model parameters. 

\vspace{1mm}\noindent\textbf{Compressing Sparse Gradient.}
Sparse-sketch consists of a bitmap and a count-sketch. The indices of non-zeros can be encoded with the bitmap. The size of the bitmap is the number of word vectors or gradient blocks, and each element of the bitmap indicate whether the block is non-zero (contains non-zero gradients). 
When several bitmaps aggregate in All-Reduce, the non-zero elements of the result bitmap indicate the non-zero blocks among them.  
We compress the value of sparse gradient with the count-sketch, which reduces the space cost by storing a large set of items over a small number of bucket arrays. Here we take such a sketch structure that contains one bucket array as an example. Let $g \in \mathbb{R}^d$ denote the vectors of gradients (including the zeros) with size $ d $ and let $m$ denote the number of buckets in sketch $ S $, respectively. Let $ s \in \{-1, +1\}^{d }$ denote a vector of random sign. In the compression phase, only the non-zero gradient $g(i)$ are randomly mapped to a bucket by a hash function $h$, adding a weighted value $ s(i)g(i) $ upon the current value of the bucket. This procedure is illustrated on the left side of Fig.~\ref{fig:sparsesketch}. The zeros are recorded with the bitmap and have no effect on the bucket value. Thus, our method ignores the zeros, the result is identical to a sketch that compresses the dense gradient.


\vspace{1mm}\noindent\textbf{Collecting Sparse-Sketch with AllReduce.}
Here, we prove that the compressed sparse gradient values can be added up directly, which is a prerequisite for AllReduce. Let $A \in \{-1, 0, +1\}^{d \times m}$ denotes the mapping matrix where $ A(i,j) = +1 $ or $ -1 $ iff the $i_{th}$ item ${g}_i$ is mapped to the $j_{th}$ bucket where $j\in [1,m]$, and to zero otherwise. Let the $j_{th} $ row vector of $ A^{\top} $ denote as $ A^{\top}(j, :) $. Then, the value of $ j_{th} $ bucket is $ A^{\top}(j,:)g  $. Consequently, the insertion phase of compressing gradient by count sketch can be mathematically represented as a stochastic projection model: $A^{\top} g$. Similarly, we query $ g(i) $ with on $ i_{th} $ row vector of $ A $ in the decompression phase: $ A(i, :)  A^{\top} g$. Therefore, the estimation values of sketch can be represented as $ AA^{\top}X $, which is a linear operation. When we have $ W $ compressed gradient $ {A}^{\top} g_0, {A}^{\top} g_1, \dots, {A}^{\top} g_{W-1} $, the aggregated gradient is $ {A}^{\top}\sum{g_k} $. Therefore, the aggregated sparse sketch is equivalent to that produced on the sum of sparse gradient.

\vspace{1mm}\noindent\textbf{Decompressing Sparse Gradient.}
The decompression phase is illustrated in the right part of Fig.~\ref{fig:sparsesketch}. As mentioned above, we precisely record the indices of non-zeros with the aggregated bitmap, it is straightforward to recover the indices of non-zeros from the aggregated bitmap. Then,  the non-zero gradient $ g(i) $ is queried as the median of the weighted values $ s(i)S[h(i)] $ in the indexed buckets. Although the queried value might not be the original value due to hash collision, we show that the estimated non-zero value is an unbiased estimation for the original gradient in Section~\ref{subsec:conver}. Let $ \alpha $ denote the percentage of non-zero elements, our method can compress the sparse data and consequently reduce the communication cost when the size of sketch $ \lambda=\frac{m}{\alpha d}<1  $.
The detail of our method is summarized in Algorithm~\ref{algo:sgdcas}.

\begin{algorithm} [!t]\small
	\caption{Sparse-Sketch Reducer}
	\label{algo:sgdcas}
	\begin{algorithmic} [1]
		\Require 
		$ Sparse(g) $ - the indices of non-zeros;
		$ r,c $ - size of sketch.
		\Ensure Aggregated sparse gradient $ \hat{g} $.
		\renewcommand{\algorithmicrequire}{\textbf{Initialize:}}  
		\Require sign hashes $ \{ s_{ij}\} $and bucket hashes$ \{ h_{ij}\} $, $ r \times c  $ table of count-sketch $ S $, bitmap $ C $.

		\ForAll{Worker $w \in [0,W-1]$}
		
		
		\State $ C[x] \leftarrow 1 $ if $ x_{th}  $ block is non-zero
		\ForAll{$i\in  Sparse(g)$}
		\ForAll{$j \in [0, r-1]$}
		\State $ S[j, h_j(i)]+=s_j(i)g(i)$ // Compression
		\EndFor
		\EndFor  
		\State AllReduce( $ S, C$) 
		
		\State Recover $ Sparse(g) $ from aggregated bitmap $ C $
		\ForAll { $ i \in Sparse(g)$}
		
		\ForAll{$j \in [0, r-1]$}
		\State $ estimates[j, i] = s_j(i)S[j, h_j(i)]$
		\EndFor
		\State $ \hat{g}(i) \leftarrow median(estimates) $ // Decompression 
		\EndFor
		\State \Return $\hat{g}$
		\EndFor
	\end{algorithmic}
\end{algorithm}

\subsection{The Convergence of S2 Reducer Based SGD}\label{subsec:conver}
The other challenge of applying S2 Reducer in distributed SGD is ensuring convergence. There are two origins of sparse gradient, one is the embedding matrix, the other is the gradient sparsification. We theoretically prove that our method maintains the convergence rate in both settings. We state the detailed proof in Appendix~\ref{appen:theo1} and ~\ref{appen:theo2} of~\cite{ge2021s2}.

Firstly, we prove that the gradient queried from sparse-sketch is an unbiased estimation of the sparse gradient. This result immediately derives in Theorem~\ref{theo:csmean} that our method has the same order of convergence rate as vanilla distributed SGD when applying to the embedding layer gradients.

\begin{theorem}\label{theo:csmean}
	Let $ \hat{g} $ denote the queried gradient. The expectation of the estimation error is zero, \textit{i.e.,} $ \mathbb{E}\left[\hat{g}\right]=g$. S2 Reducer Based SGD updates model parameters $\boldsymbol{\omega}^t\in {\mathbb{R}^d}$ with the gradients of  loss function  ${f}(\boldsymbol{\omega}_{t})$ at iteration $t$, the convergence rate is $ \min\limits_{0\leq t\leq{T}} {\left \| \nabla{f(\boldsymbol{\omega}^t)} \right \|}^2=O(1/\sqrt{T}) $. 
\end{theorem} 

Next, we prove the convergence of the S2 Reducer-based SGD matches the convergence rate of vanilla block Top-K method in Theorem~\ref{theo:error}. 
We derive that our scheme is a $ \delta $-approximate compressor. Then the convergence result is given based on the proof of~\cite{karimireddy2019error}.



\begin{definition}[$\delta$-approximate compressor]{\label{def:delta}}
	An operator $\mathcal{U}(\cdot):  \mathbb{R}^d \rightarrow \mathbb{R}^d $ is a $ \delta $-approximate compressor if $\exists \delta \in (0, 1) $ such that
	$\mathbb{E}\left\| g- \mathcal{U}(g)  \right\|^2_2 \leq  \left(1-\delta\right)\left\| g \right\|^2_2 $, $\forall  g  \in \mathbb{R}^d $.
\end{definition}

\begin{theorem}\label{theo:error}
	Let $ b $ denote the total number of blocks. The sketch based block Top-K are $ \delta $-approximate compressors with  $ \delta = \frac{K}{b} $, \textit{i.e.,} $ \mathbb{E}\left\| g-\mathcal{U}({sparse}(g))\right\|_{2}^{2}= \left(1-\frac{K}{b}\right)\left\| g\right\|_{2}^{2}$. The convergence rate is $ \min\limits_{0\leq t\leq{T}} {\left \| \nabla{f(\boldsymbol{\omega}^t)} \right \|}^2=O(1/\sqrt{T}) $.
	
\end{theorem}

\section{Evaluation}\label{sec:exp}

\begin{figure}[th]
	\renewcommand{\arraystretch}{1.0}
	\centering
	\begin{subfigure}[b]{0.45\linewidth}
		\includegraphics[width=\linewidth]{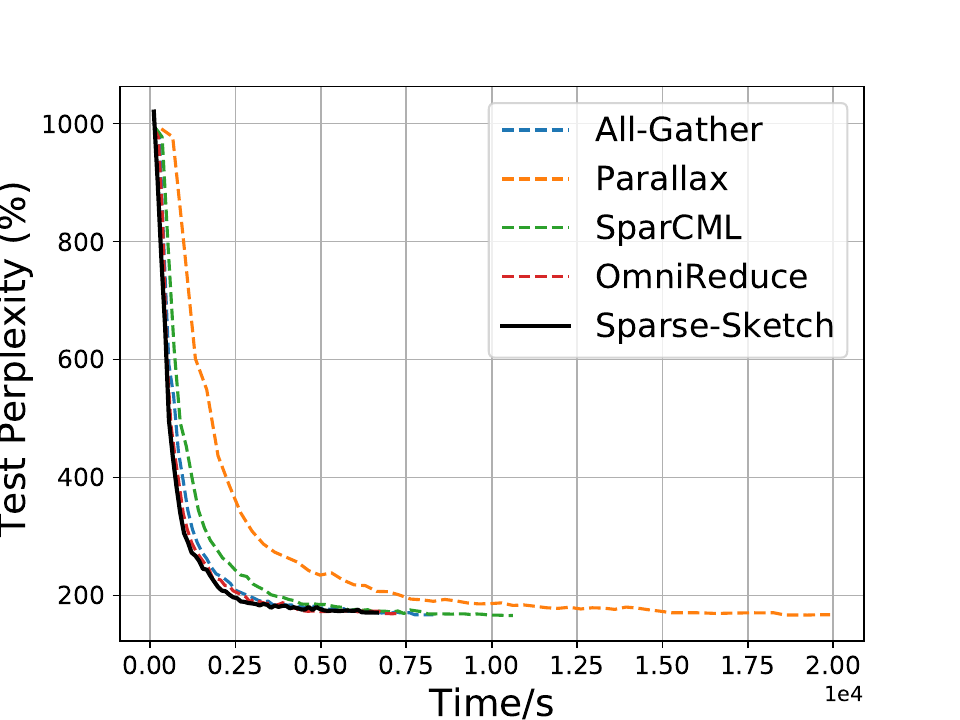}
		\caption{LSTM}
	\end{subfigure}
	\begin{subfigure}[b]{0.45\linewidth}
		\includegraphics[width=\linewidth]{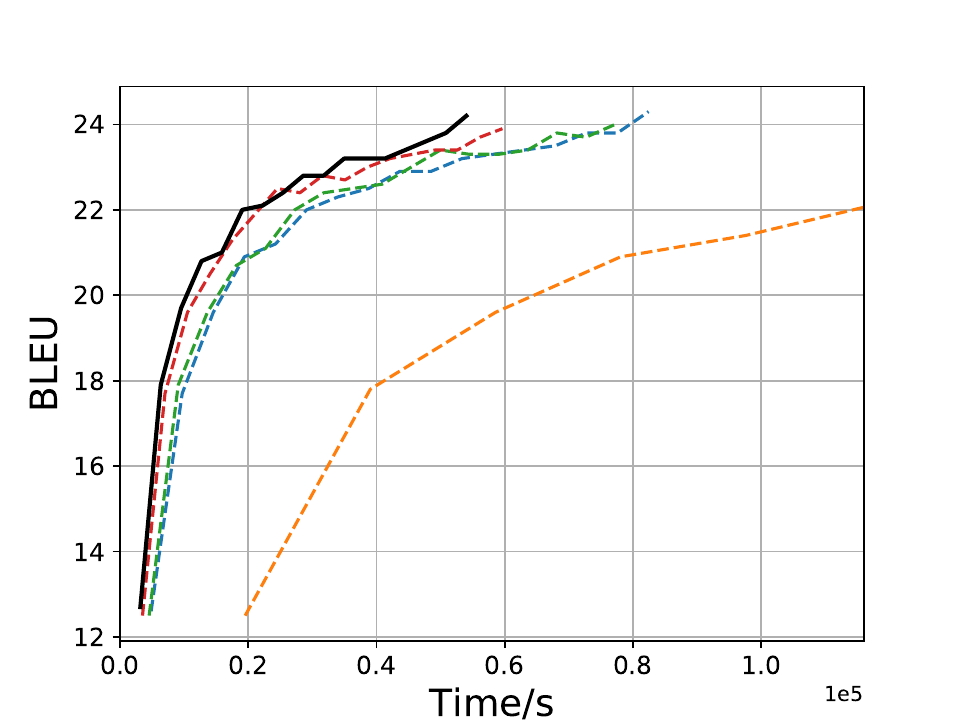}
		\caption{GNMT-8}
	\end{subfigure}
	\begin{subfigure}[b]{0.45\linewidth}
		\includegraphics[width=\linewidth]{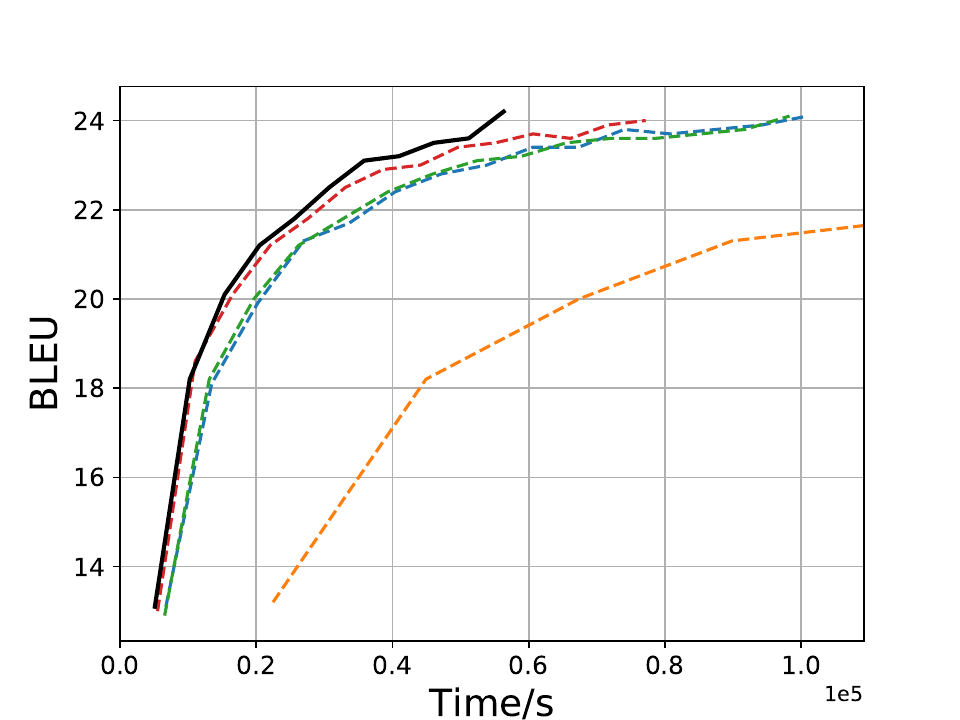}
		\caption{GNMT-16}
	\end{subfigure}
	\begin{subfigure}[b]{0.45\linewidth}
	\includegraphics[width=\linewidth]{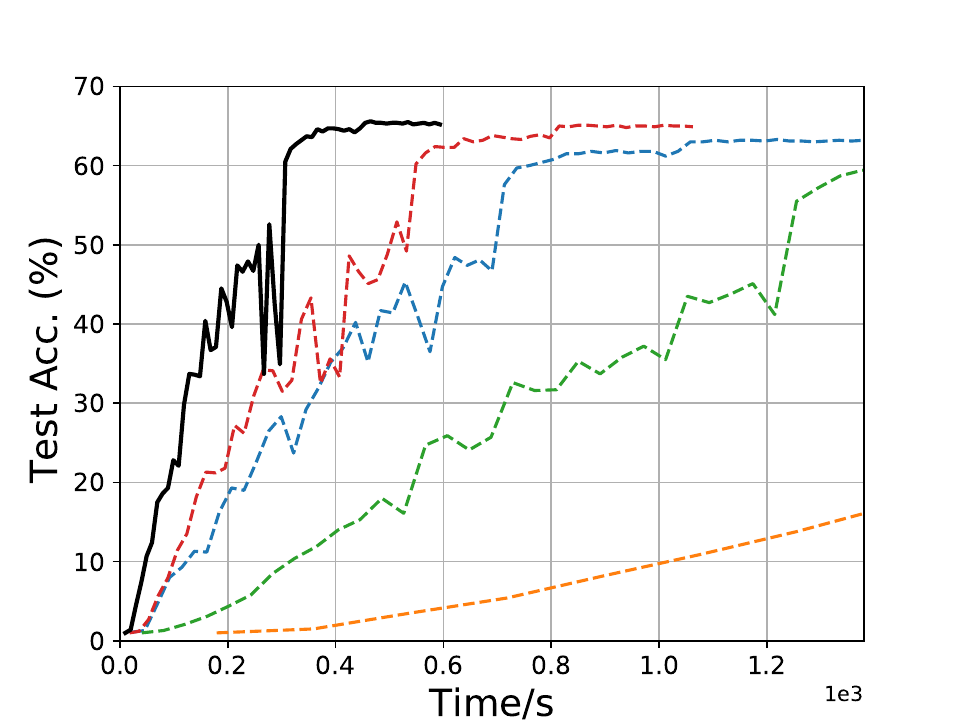}
	\caption{Sparse ResNet50}
	\end{subfigure}
	\begin{subfigure}[b]{0.45\linewidth}
		\includegraphics[width=\linewidth]{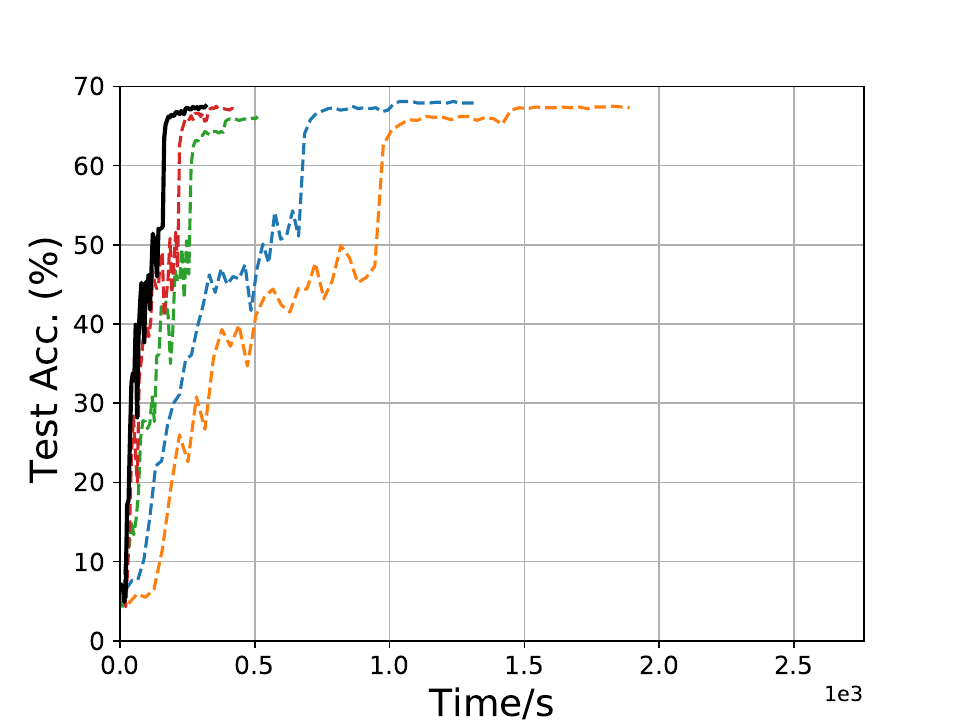}
		\caption{Sparse DenseNet}
	\end{subfigure}
	\begin{subfigure}[b]{0.45\linewidth}
		\includegraphics[width=\linewidth]{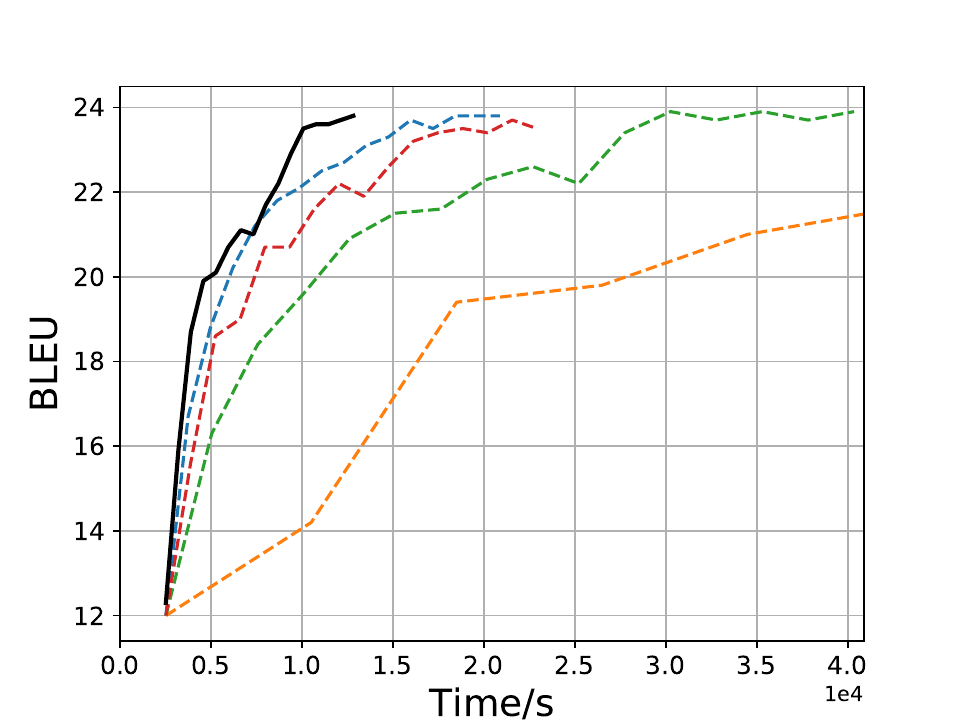}
		\caption{Sparse GNMT-8}
	\end{subfigure} 
	\caption[Validation loss]{Training curves of (a)-(c): NLP models, where the embedding layers gradients are sparse. (d)-(f): block-wise sparsification.}
	\label{fig:test_loss_embedding}
\end{figure}

In this section, we show the end-to-end training speedup of the S2 Reducer compared with four baselines: All-Gather, Parallax~\cite{kim2019parallax}, SparCML~\cite{renggli2019sparcml}, OmniReduce~\cite{fei2021efficient}. We have implemented these methods based on PyTorch, and deployed them on a 16 RTX-3090 GPU cluster that equips a 10Gb Ethernet. The communication library is NCCL~\cite{jeaugey2017nccl} except Parallax, which can only use Gloo.

We conduct two kinds of experiments. Firstly, we demonstrate the speedup when training LSTM for language models, GNMT-8 and GNMT-16~\cite{wu2016google}, which yield sparse gradients from embedding layers. Then we show the performance of accelerating the training of ResNet-50~\cite{he2016deep} DenseNet~\cite{huang2017densely} and GNMT-8 when sparsifying the whole gradients. The details of models and datasets are shown in Table~\ref{tab:model_info}.

\begin{figure}[t]
	\centering
	\includegraphics[width=0.7\linewidth]{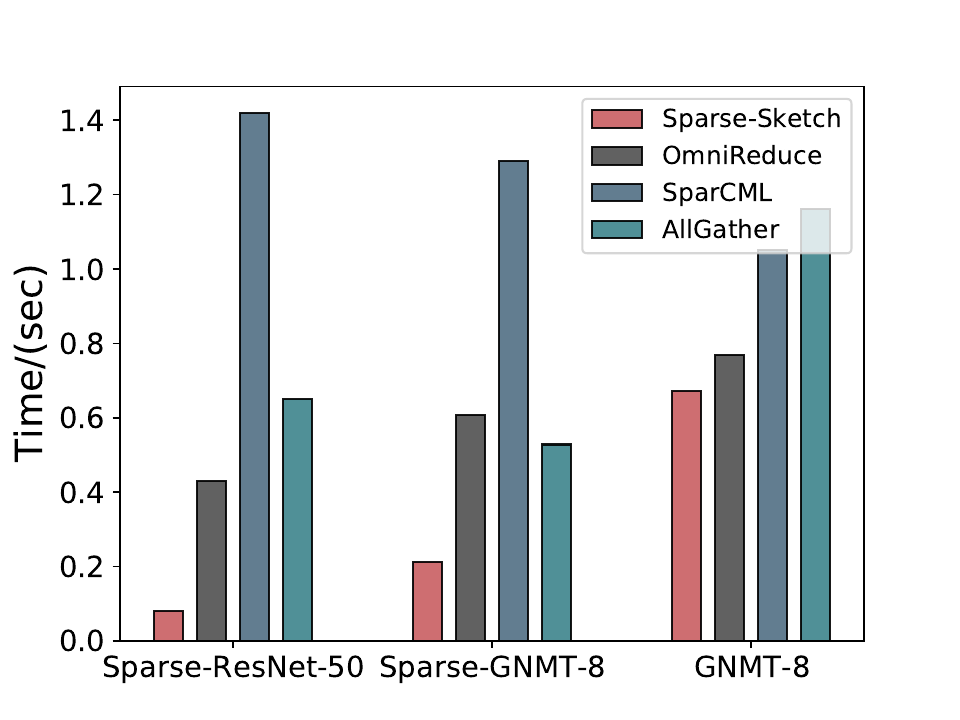}
	\caption{The communication overhead of sparse gradient. The first two models are ResNet-50 and GNMT-8 with block-wise sparsification, and the last one yields parse gradient from embedding layers.}
	\label{fig:sparsecomm_time}
\end{figure}

\begin{table}[]
	\renewcommand{\arraystretch}{0.9}
	\centering
	\begin{tabular}{|c|c|c|c|}
		\hline
		\textbf{Model}     & \textbf{Dataset}    & \textbf{Embed./Dense}& \textbf{Sparsity} \\ \hline
		LSTM      & WikiText-2& 75/115 MB       & 99.1\%            \\ \hline
		GNMT-8    & WMT-16     & 265/195 MB       & 95.17\%            \\ \hline
		GNMT-16   & WMT-16     & 265/245 MB       & 95.17\%            \\ \hline
		DenseNet  & CIFAR-100 & 4.04 MB      &$ -- $         \\ \hline
		ResNet-50 & CIFAR-100  & 94.8 MB      &$ -- $           \\ \hline
	\end{tabular}
	\caption{The details of training models and datasets. The first three models contain embedding layers, the table lists the size of embedding and other dense weights. We represent the sparsity by the average percentage of non-zeros in embedding layer gradient. The last two are dense models, we apply block-wise sparsification on them. }
	\label{tab:model_info}
\end{table}

\vspace{1mm}\noindent\textbf{Sparse Gradient from Embedding Layers.}
As illustrated in Fig.~\ref{fig:test_loss_embedding} (a) \& (b), our method achieves $1.6\times$ speedup compared with SparCML when training LSTM, and also $ 1.25\times $ speedup compared with AllGather implementation on GNMT-8. Although our method has compressed the gradient, the model accuracy not only is the same as other lossless methods when training the same epochs in LSTM and GNMT-8 but also converges faster to the target BLEU on GNMT-16 than others.  Fig.~\ref{fig:test_loss_embedding} (c) show that our method achieves $ 1.4\times $ speedup compares with OmniReduce, the state-of-the-art method, with only 11 epochs, while others require 16 epochs.

\vspace{1mm}\noindent\textbf{Block-wise Sparse Gradient.}
In this section, we apply block-wise sparsification on DenseNet, ResNet-50, and GNMT-8 with a $ 1/32 $ compression rate. Due to the small size of DenseNet and ResNet-50 after sparsification, we configure the bandwidth to 1GbE in their training. As illustrated in Fig.~\ref{fig:test_loss_embedding} (f), our method achieves $ 1.8\times $ speedup compared with OmniReduce, which is the SOTA approach, on the large NLP model GNMT-8, and reduce 65\% sparse communication time. As shown in Fig.~\ref{fig:test_loss_embedding} (d), our method reduces 81\% sparse communication time compared with OmniReduce, and achieves $ 4.1\times $ speedup over SparCML and $ 2.3\times $ over AllGather on ResNet-50 (block-wise sparsification), while maintaining the model accuracy. 
 The detailed results of communication overhead and sensitivity of hyper-parameter $ \lambda $ are stated in Appendix~\ref{appen:sparse_time} and~\ref{appen:sensitivity} of~\cite{ge2021s2} respectively.

\section{Conclusion}
This paper proposes Sparse-Sketch Reducer, a novel sketch-based sparse gradient collective communication method. Our method reduces the communication overhead by enabling AllReduce on the compressed sparse-sketch. Experimental results show that it effectively reduces communication costs and accelerates distributed DNN training. 

\bibliographystyle{IEEEbib}
\small\balance \bibliography{IEEEabrv,IEEEexample}

\begin{thebibliography}{10}

\bibitem{he2016deep}
Kaiming He, Xiangyu Zhang, Shaoqing Ren, and Jian Sun,
\newblock ``Deep residual learning for image recognition,''
\newblock in {\em CVPR}, 2016, pp. 770--778.

\bibitem{huang2017densely}
Gao Huang, Zhuang Liu, Laurens Van Der~Maaten, and Kilian~Q Weinberger,
\newblock ``Densely connected convolutional networks,''
\newblock in {\em CVPR}, 2017, pp. 4700--4708.

\bibitem{dosovitskiy2020image}
Alexey Dosovitskiy, Lucas Beyer, Alexander Kolesnikov, Dirk Weissenborn,
  Xiaohua Zhai, Thomas Unterthiner, Mostafa Dehghani, Matthias Minderer, Georg
  Heigold, Sylvain Gelly, et~al.,
\newblock ``An image is worth 16x16 words: Transformers for image recognition
  at scale,''
\newblock in {\em ICLR}, 2020.

\bibitem{gulati2020conformer}
Anmol Gulati, James Qin, Chung-Cheng Chiu, Niki Parmar, Yu~Zhang, Jiahui Yu,
  Wei Han, Shibo Wang, Zhengdong Zhang, Yonghui Wu, et~al.,
\newblock ``Conformer: Convolution-augmented transformer for speech
  recognition,''
\newblock {\em arXiv preprint arXiv:2005.08100}, 2020.

\bibitem{xu2021self}
Qiantong Xu, Alexei Baevski, Tatiana Likhomanenko, Paden Tomasello, Alexis
  Conneau, Ronan Collobert, Gabriel Synnaeve, and Michael Auli,
\newblock ``Self-training and pre-training are complementary for speech
  recognition,''
\newblock in {\em ICASSP 2021}. IEEE, 2021, pp. 3030--3034.

\bibitem{dai2019transformer}
Zihang Dai, Zhilin Yang, Yiming Yang, Jaime~G Carbonell, Quoc Le, and Ruslan
  Salakhutdinov,
\newblock ``Transformer-xl: Attentive language models beyond a fixed-length
  context,''
\newblock in {\em ACL}, 2019, pp. 2978--2988.

\bibitem{mehta2020delight}
Sachin Mehta, Marjan Ghazvininejad, Srinivasan Iyer, Luke Zettlemoyer, and
  Hannaneh Hajishirzi,
\newblock ``Delight: Deep and light-weight transformer,''
\newblock in {\em ICLR}, 2020.

\bibitem{ben2019demystifying}
Tal Ben-Nun and Torsten Hoefler,
\newblock ``Demystifying parallel and distributed deep learning: An in-depth
  concurrency analysis,''
\newblock {\em ACM CSUR}, vol. 52, no. 4, pp. 1--43, 2019.

\bibitem{li2014scaling}
Mu~Li, David~G Andersen, Jun~Woo Park, Alexander~J Smola, Amr Ahmed, Vanja
  Josifovski, James Long, Eugene~J Shekita, and Bor-Yiing Su,
\newblock ``Scaling distributed machine learning with the parameter server,''
\newblock in {\em $\{$USENIX$\}$ OSDI 14}, 2014, pp. 583--598.

\bibitem{fan2021dapple}
Shiqing Fan, Yi~Rong, Chen Meng, Zongyan Cao, Siyu Wang, Zhen Zheng, Chuan Wu,
  Guoping Long, Jun Yang, Lixue Xia, et~al.,
\newblock ``Dapple: A pipelined data parallel approach for training large
  models,''
\newblock in {\em ACM SIGPLAN}, 2021, pp. 431--445.

\bibitem{abadi2016tensorflow}
Mart{\'\i}n Abadi, Paul Barham, Jianmin Chen, Zhifeng Chen, Andy Davis, Jeffrey
  Dean, Matthieu Devin, Sanjay Ghemawat, Geoffrey Irving, Michael Isard,
  et~al.,
\newblock ``Tensorflow: A system for large-scale machine learning,''
\newblock in {\em $\{$USENIX$\}$ OSDI 16}, 2016, pp. 265--283.

\bibitem{paszke2019pytorch}
Adam Paszke, Sam Gross, Francisco Massa, Adam Lerer, James Bradbury, Gregory
  Chanan, Trevor Killeen, Zeming Lin, Natalia Gimelshein, Luca Antiga, et~al.,
\newblock ``Pytorch: An imperative style, high-performance deep learning
  library,''
\newblock in {\em NeurIPS}, 2019, pp. 8026--8037.

\bibitem{rasley2020deepspeed}
Jeff Rasley, Samyam Rajbhandari, Olatunji Ruwase, and Yuxiong He,
\newblock ``Deepspeed: System optimizations enable training deep learning
  models with over 100 billion parameters,''
\newblock in {\em ACM SIGKDD}, 2020, pp. 3505--3506.

\bibitem{alistarh2017qsgd}
Dan Alistarh, Demjan Grubic, Jerry Li, Ryota Tomioka, and Milan Vojnovic,
\newblock ``Qsgd: Communication-efficient sgd via gradient quantization and
  encoding,''
\newblock in {\em NeurIPS}, 2017, pp. 1709--1720.

\bibitem{lin2018deep}
Yujun Lin, Song Han, Huizi Mao, Yu~Wang, and Bill Dally,
\newblock ``Deep gradient compression: Reducing the communication bandwidth for
  distributed training,''
\newblock in {\em ICLR}, 2018.

\bibitem{jeaugey2017nccl}
Sylvain Jeaugey,
\newblock ``Nccl 2.0,''
\newblock in {\em GTC}, 2017.

\bibitem{peng2019generic}
Yanghua Peng, Yibo Zhu, Yangrui Chen, Yixin Bao, Bairen Yi, Chang Lan, Chuan
  Wu, and Chuanxiong Guo,
\newblock ``A generic communication scheduler for distributed dnn training
  acceleration,''
\newblock in {\em SOSP}, 2019, pp. 16--29.

\bibitem{cho2019blueconnect}
Minsik Cho, Ulrich Finkler, and David Kung,
\newblock ``Blueconnect: Novel hierarchical all-reduce on multi-tired network
  for deep learning,''
\newblock in {\em Proceedings of the 2nd SysML Conference}, 2019.

\bibitem{fei2021efficient}
Jiawei Fei, Chen-Yu Ho, Atal~N Sahu, Marco Canini, and Amedeo Sapio,
\newblock ``Efficient sparse collective communication and its application to
  accelerate distributed deep learning,''
\newblock in {\em SIGCOMM 2021}, 2021, pp. 676--691.

\bibitem{vogels2019powersgd}
Thijs Vogels, Sai~Praneeth Karimireddy, and Martin Jaggi,
\newblock ``Powersgd: Practical low-rank gradient compression for distributed
  optimization,''
\newblock in {\em NeurIPS}, 2019, pp. 14259--14268.

\bibitem{kim2019parallax}
Soojeong Kim, Gyeong-In Yu, Hojin Park, Sungwoo Cho, Eunji Jeong, Hyeonmin Ha,
  Sanha Lee, Joo~Seong Jeong, and Byung-Gon Chun,
\newblock ``Parallax: Sparsity-aware data parallel training of deep neural
  networks,''
\newblock in {\em EuroSys 2019}, 2019, pp. 1--15.

\bibitem{renggli2019sparcml}
C{\`e}dric Renggli, Saleh Ashkboos, Mehdi Aghagolzadeh, Dan Alistarh, and
  Torsten Hoefler,
\newblock ``Sparcml: High-performance sparse communication for machine
  learning,''
\newblock in {\em SC}, 2019, pp. 1--15.

\bibitem{ivkin2019communication}
Nikita Ivkin, Daniel Rothchild, Enayat Ullah, Ion Stoica, Raman Arora, et~al.,
\newblock ``Communication-efficient distributed sgd with sketching,''
\newblock in {\em NeurIPS}, 2019, pp. 13144--13154.

\bibitem{agarwal2013mergeable}
Pankaj~K Agarwal, Graham Cormode, Zengfeng Huang, Jeff~M Phillips, Zhewei Wei,
  and Ke~Yi,
\newblock ``Mergeable summaries,''
\newblock {\em ACM TODS}, vol. 38, no. 4, pp. 1--28, 2013.

\bibitem{ge2021s2}
Keshi Ge, Yongquan Fu, Zhiquan Lai, Xiaoge Deng, and Dongsheng Li,
\newblock ``S2 reducer: High-performance sparse communication to accelerate
  distributed deep learning,'' 2021.

\bibitem{karimireddy2019error}
Sai~Praneeth Karimireddy, Quentin Rebjock, Sebastian Stich, and Martin Jaggi,
\newblock ``Error feedback fixes signsgd and other gradient compression
  schemes,''
\newblock in {\em ICML}. PMLR, 2019, pp. 3252--3261.

\bibitem{wu2016google}
Yonghui Wu, Mike Schuster, Zhifeng Chen, Quoc~V Le, Mohammad Norouzi, Wolfgang
  Macherey, Maxim Krikun, Yuan Cao, Qin Gao, Klaus Macherey, et~al.,
\newblock ``Google's neural machine translation system: Bridging the gap
  between human and machine translation,''
\newblock {\em arXiv preprint arXiv:1609.08144}, 2016.

\end{thebibliography}

\onecolumn
\appendix

\section{Discussion of Convergence}\label{sec:appen1}

\begin{assumption}{\label{assum:smooth}}
	In distributed setting with $ W $ worker nodes, let function $ f: \mathbb{R}^d \rightarrow \mathbb{R}$ is L-smooth, and for any $ \boldsymbol{\omega} \in \mathbb{R}^d $, both the local and global gradients are bounded, i.e.: $\exists \sigma_{i}, \sigma \in \mathbb{R} $ such that $ \left\|\nabla f_{i}({\boldsymbol{\omega}})\right\| \leq \sigma_{i},\|\nabla f(\boldsymbol{\omega})\| \leq \sigma $.
\end{assumption}

\subsection{Proof of Theorem~\ref{theo:csmean}}{\label{appen:theo1}}

\begin{proof}
	The error of a sketch can be quantified as $\hat{g}-g=A A^{T} g-g=\left(A A^{T}-I\right) g$, where $I$ denotes the identity matrix. $AA^T$ is a $d\times d$ symmetric matrix, where each entry $(i, j)$ is one when the $i^{th}$ and $j^{th}$ gradients mapped into the same bucket, {i.e.}: $A(i, :)=A(j, :)$, and zero otherwise. It means that the diagonal entries are all set to ones.  Each gradients is mapped in a uniformly-random manner, so that $Pr[AA^T(i, j)=+1]=Pr[AA^T(i, j)=-1]=\frac{0.5}{m}$ and $Pr[AA^T(i, j)=0]=1-\frac{1}{m}$, which follows Bernoulli distribution. 
	Let $\Phi=AA^T-I$, then each non-diagonal entry in $\Phi$ is the same as that in $AA^T$, and the diagonal entries are zeros. From the probability distribution, we have $\mathbb{E}[\Phi(i, j)]=0$.
	
	Thus, We can deduce the expectation of estimation error respect to ${g}(j)$ as below: 
	\begin{align*}\label{eq:csmean}
		\mathbb{E}\left[ \hat{g}(j)-g(j)\right] 
		&=\mathbb{E}\left[\sum_{i=1}^{d-1} \Phi_{i\neq j}(i, j) g(j)\right] \\
		&=\sum_{i=1}^{d-1} \mathbb{E}\left[\Phi_{i\neq j}(i, j) g(j)\right] \\ 
		&=\sum_{i=1}^{d-1} \mathbb{E}\left[\Phi_{i\neq j}(i, j)\right] \mathbb{E}[g(j)]=0
	\end{align*}
	
	Using the $L$-smoothness  property of the loss function $f$ (Assumption~\ref{assum:smooth}) and the fact that $ \mathbb{E}\left[\hat{g}\right]=g$, we have that
	\begin{equation}
		\label{eqn_39}
		\begin{split}
			\mathbb{E}[f(w_{t+1})]-\mathbb{E}[f(w_{t})] &\leq \mathbb{E}\langle \nabla f(w_{t}), w_{t+1}-w_{t} \rangle +\frac{L}{2} \mathbb{E}\|w_{t+1}-w_{t}\|_{2}^{2}\\
			&\leq \langle \nabla f({\omega}_{t}), w_{t+1}-w_{t} \rangle + \frac{L \eta^{2}}{2}\|g_{t}\|_{2}^{2}\\
			&\leq -\eta \|\nabla f({\omega}_{t})\|_{2}^{2}+\frac{L \eta^{2} \sigma^{2}}{2},
		\end{split}
	\end{equation}
	
	where $\eta$ is the learning rate and $\mathbb{E}\|g\|_{2}^{2} \leq \sigma^{2}$.
	
	Now rearranging the terms and averaging over $T$ gives
	\begin{equation}
		\begin{aligned}
			\frac{1}{T+1} \sum_{t=0}^{T}\left\|\nabla f\left(w_{t}\right)\right\|_{2}^{2} & \leq \frac{1}{\eta(T+1)} \sum_{t=0}^{T}\left(\mathbb{E}\left[f\left(w_{t}\right)\right]-\mathbb{E}\left[f\left(w_{t+1}\right)\right]\right)+\frac{L \eta^{2} \sigma^{2}}{2} \\
			& \leq \frac{f\left(w_{0}\right)-f^{\star}}{\eta(T+1)}+\frac{L \eta^{2}\sigma^{2}}{2},
		\end{aligned}
	\end{equation}
	
	where $f^{*}$ is the optimal value. With the decreasing learning rate $\eta=\frac{1}{\sqrt{T}}$, we have
	$$ \min\limits_{0\leq t\leq{T}} {\left \| \nabla{f(\boldsymbol{\omega}_t)} \right \|}_{2}^2=O(\frac{1}{\sqrt{T}}). $$
\end{proof}

\subsection{Proof of Theorem~\ref{theo:error}}{\label{appen:theo2}}

\begin{proof}
	Let $\mathcal{U}(\cdot)$ denote the unbiased count sketch compressor, $ {sparse}(g) $ denote the non-zero gradient after block Top-K sparsification. Let $ g_i\in \mathbb{R}^{\frac{d}{b}}  $ denote the $  i_{th} $ block for any $ g \in \mathbb{R}^d $.
	Let $ S_{block-topk}$ denote the set of Top-K blocks corresponding to a given $ g $. Then, 
	
	\begin{align*}
		\mathbb{E}\left\| \mathcal{U}({sparse}(g)) \right\|_{2} ^{2}
		&=\left\| {sparse}(g) \right\|_{2} ^{2} \\ 
		&=\sum_{i \in S_{block-topk}}  \left\|g_{i}\right\|_{2}^{2} \\ 
		&=\sum_{i \in S_{block-topk}} \sum_{k=1}^{{d}/{b}}\left\|g_{i}^{k}\right\|_{2}^{2} \\ 
		& \geq \frac{K}{b}\|g\|_{2} ^{2}
	\end{align*}
	
	Where $ g_{i}^{k} $ denotes the $ k_{th} $ element of $ i_{th} $ block.
	
	Hence, we have
	
	\begin{align*} 
		\mathbb{E}\left\| g-\mathcal{U}({sparse}(g))\right\|_{2}^{2}
		=&\left\| g\right\|_{2}^{2}+ \mathbb{E} \left\| \mathcal{U}({sparse}(g)) \right\|_{2}^{2} \\
		&-2\mathbb{E}\left\langle g, \mathcal{U}({sparse}(g)) \right\rangle \\
		=&\left\| g\right\|_{2}^{2}- \left\| {sparse}(g) \right\|_{2}^{2} \\ 
		\leq &\left\| g\right\|_{2}^{2}-\frac{K}{b}\left\| g\right\|_{2}^{2} \\
		= & \left(1-\frac{K}{b}\right)\left\| g\right\|_{2}^{2}
	\end{align*}
	
	That means $\mathcal{U}(\cdot)$ is a $\delta$-approximate compressor. Combined with the definition of the error sequence, we have
	\begin{align*}
		\left\|{e}_{t+1}^{i}\right\|_{2}^{2}=&\left\|{g}_{t}^{i}-\mathcal{U}({g}_{t}^{i})\right\|_{2}^{2}\leq(1-\delta)\|{g}_{t}^{i}\|_{2}^{2}\\
		=&(1-\delta)\|\nabla f_{i}({\omega}_{t})+{e}_{t}^{i}\|_{2}^{2}\\
		\le& (1-\delta)\big[(1+\gamma)\|{e}_{t}^{i}\|_{2}^{2}+(1+1/\gamma)\|\nabla f_{i}({\omega}_{t})\|_{2}^{2}\big]\\
		\le&(1-\delta)(1+\gamma)^{t+1}\|{e}_{0}^{i}\|_{2}^{2}+\sum_{k=0}^{t}\big[(1-\delta)(1+\gamma)\big]^{t-k}(1-\delta)(1+1/\gamma)\|\nabla f_{i}({\omega}_{t})\|_{2}^{2}\\
		\le& \sum_{k=0}^{\infty}\big[(1-\delta)(1+\gamma)\big]^{t-k}(1-\delta)(1+1/\gamma) \sigma_{i}^{2}\\
		=&\frac{(1-\delta)(1+1 / \gamma)}{1-(1-\delta)(1+\gamma)} \sigma_{i}^{2}.
	\end{align*}
	
	Let us pick $\gamma=\frac{\delta}{2(1-\delta)}$ such that $1+1/\gamma=(2-\delta)/\delta \leq 2/\delta$. Then we have
	\begin{align*}
		\left\|{e}_{t+1}^{i}\right\|_{2}^{2}\le\frac{2(1-\delta)(1+1 / \gamma)}{\delta} \sigma_{i}^{2} \leq \frac{4(1-\delta)}{\delta^{2}} \sigma_{i}^{2}.
	\end{align*}
	
	We define an auxiliary sequence $\{{\nu}_{t}\}_{t=0,1,\cdots}$, which has the following property
	\begin{align*}
		{\nu}_{t+1}&={\omega}_{t+1}-\eta  \sum_{i=1}^{W}\mathbf{e}_{t+1}^{i}={\nu}_{t}-\eta \nabla f({\omega}_{t}).
	\end{align*}
	
	Since the function $f$ is $L$-smooth (Assumption~\ref{assum:smooth}), We have
	
	\begin{equation}
		\begin{split}
			\mathbb{E}f({\nu}_{t+1})-\mathbb{E}f({\nu}_{t}) \leq& \langle \nabla f({\nu}_{t}), {\nu}_{t+1}-{\nu}_{t} \rangle +\frac{L}{2}\|{\nu}_{t+1}-{\nu}_{t}\|_{2}^{2}\\
			\leq& \langle \nabla f({\omega}_{t}), {\nu}_{t+1}-{\nu}_{t} \rangle + \langle \nabla f({\nu}_{t})-f({\omega}_{t}), {\nu}_{t+1}-{\nu}_{t} \rangle+\frac{L}{2}\|{\nu}_{t+1}-{\nu}_{t}\|_{2}^{2}\\
			\leq& -\eta(1-\frac{L\eta}{2})\|\nabla f({\omega}_{t})\|_{2}^{2}+\frac{\rho}{2}\|{\nu}_{t+1}-{\nu}_{t}\|_{2}^{2}+\frac{1}{2\rho}\| \nabla f({\nu}_{t})-f({\omega}_{t})\|_{2}^{2}\\
			\leq& -\eta \big(1-\frac{(L+\rho)\eta}{2}\big)\|\nabla f({\omega}_{t})\|_{2}^{2}+\frac{L^{2}}{2\rho}\|{\nu}_{t}-{\omega}_{t}\|_{2}^{2}\\
			\leq& -\eta\big(1-\frac{(\rho+L)\eta}{2}\big)\left\|\nabla f({\omega}_{t})\right\|_{2}^{2}+\frac{L^{2} \eta^{2}}{2 \rho}\Big\|\sum_{i=1}^{W}\mathbf{e}^{i}_{t}\Big\|_{2}^{2}.
		\end{split}
	\end{equation}
	
	The third inequality follows from the mean-value inequality and holds for any $\rho>0$. Summing the terms in (\ref{eqn_39}) over $t$ gives
	\begin{equation}
		\label{eqn_40}
		\begin{split}
			\mathbb{E}f({\nu}_{T+1})-\mathbb{E}f({\nu}_{0}) &\leq -\eta\big(1-\frac{(\rho+L)\eta}{2}\big)\sum_{t=0}^{T}\left\|\nabla f({\omega}_{t})\right\|_{2}^{2}+\frac{2L^{2} \eta^{2}(1-\delta)}{\rho \delta^{2}}(T+1)W \sigma^{2},
		\end{split}
	\end{equation}
	
	where we used (with  $\sigma^{2}:=\sum_{i=1}^{W}\sigma_{i}^{2}$)
	\begin{align*}
		\nonumber
		\|\sum_{i=1}^{W}\mathbf{e}_{t}^{i}\|_{2}^{2} &\leq W \sum_{i=1}^{W}\|\mathbf{e}_{t}^{i}\|_{2}^{2} \leq W\cdot \frac{4(1-\delta)}{\delta^{2}} \sum_{i=1}^{W}\sigma_{i}^{2}:=\frac{4(1-\delta)}{\delta^{2}} W\sigma^{2}.
	\end{align*}
	
	Noticed that ${\nu}_{0}={\omega}_{0}$ and $f^{*}$ is the optimal value. Rearranging the terms in (\ref{eqn_40}) and averaging over $t$ lead to
	\begin{align*}
		\nonumber
		\begin{split}
			\frac{1}{T+1}\sum_{t=0}^{T}\left\|\nabla f({\omega}_{t})\right\|_{2}^{2} &\leq \frac{f({\nu}_{0})-f({\nu}_{T+1})}{\eta(T+1)\big(1-\frac{(\rho+L)\eta}{2}\big)}+\frac{2L^{2} \eta(1-\delta)W \sigma^{2}}{\rho \delta^{2}\big(1-\frac{(\rho+L)\eta}{2}\big)}\\
			&\leq \frac{2(f^{0}-f^{*})}{\eta(T+1)\big(2-(\rho+L)\eta\big)}+\frac{4L^{2} \eta(1-\delta)W \sigma^{2}}{\rho \delta^{2}\big(2-(\rho+L)\eta\big)}
		\end{split}
	\end{align*}
	
	With the decreasing learning rate $\eta=\frac{1}{\sqrt{T}}$, we have
	$$ \min\limits_{0\leq t\leq{T}} {\left \| \nabla{f(\boldsymbol{\omega}_t)} \right \|}_{2}^2=O(\frac{1}{\sqrt{T}}). $$
	
\end{proof}

\section{Experiment Results \& Details}

\subsection{Collective Communication Comparison}\label{appen:allreduce_vs_allgather}
\begin{figure}[h]
	\centering
	
	\includegraphics[width=0.5\linewidth]{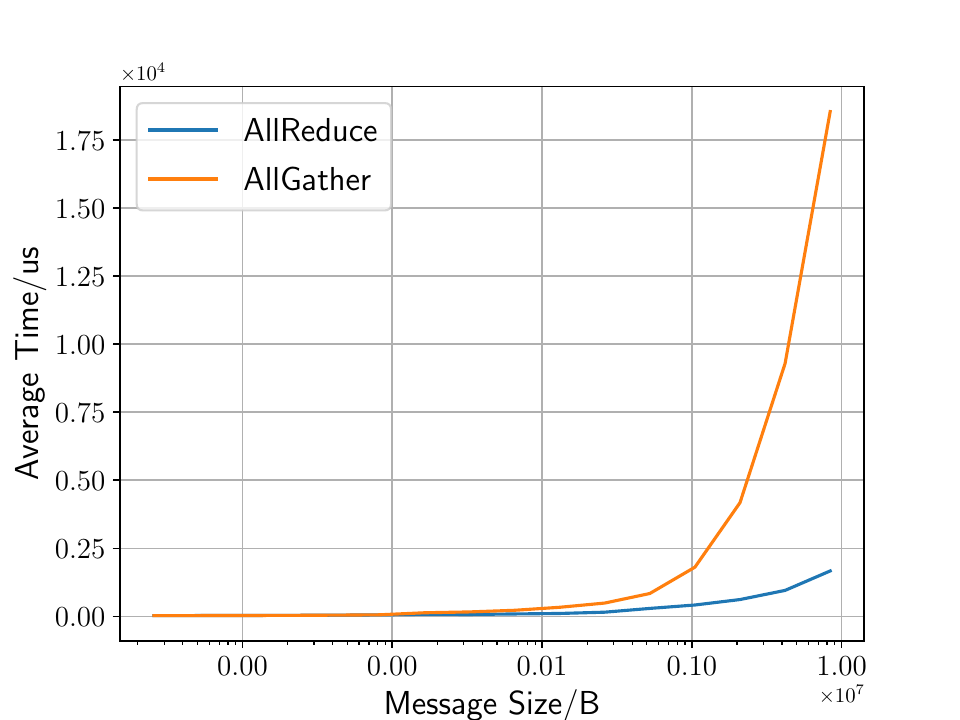}
	\caption{Collective comm. overhead on 16 GPUs}

	\caption{The comparison of the communication time between AllReduce and AllGather when synchronizing a message on a 16-GPU cluster.}
	
	\label{fig:AR_AG}
\end{figure}

We illustrate the time cost of AllReduce and AllGather when they synchronize a message. The test conduct on a 16-GPU cluster equipped with InfiniBand, the communication backend is NCCL. It shows that AllReduce is over $ 11\times $ faster than AllGather when synchronizing a data more than 8MB.  
\subsection{Sparse Gradient Communication Overhead}{\label{appen:sparse_time}}

\begin{table}[th]
	\centering
	\begin{tabular}{|c|c|c|c|c|}
		\hline
		\textbf{Model architecture} & \textbf{ResNet-50}      & \textbf{DenseNet}        & \textbf{LSTM}                 & \textbf{GNMT}       \\ \hline
		Dataset                     & \multicolumn{2}{c|}{CIFAR-100}                       & WikiText-2                      & WMT-16 German-English \\ \hline
		Number of workers & \multicolumn{4}{c|}{16}                                \\ \hline
		Hardware                    & \multicolumn{4}{c|}{4 nodes, 4 $\times$ NVIDIA GeForce RTX-3090 GPU with 24GB memory in each node}             \\ \hline
		Network           & \multicolumn{2}{c|}{1GbE} & \multicolumn{2}{c|}{10GbE} \\ \hline
		Mini-batch size   & \multicolumn{2}{c|}{128}  & 16        & 64 (128 in block-wise Top-K)             \\ \hline
		Epoch             & \multicolumn{3}{c|}{60}               & 16             \\ \hline
		Learning rate     & \multicolumn{3}{c|}{0.1 × 16 for SGD} & $1e^{-3}$      \\ \hline
		LR decay                    & \multicolumn{2}{c|}{$\times$ 0.1 at epoch 30 and 45} & $\times$ 0.1 at epoch 45 and 55 & None                  \\ \hline
	\end{tabular}
	\caption{Training setting for the experiments in Section and Appendix}
	\label{tab:exp_setting}
\end{table}

Section~\ref{sec:exp} shows the end-to-end performance of our method on the custom implementations of ResNet-50\footnotemark, DenseNet\footnotemark[\value{footnote}]\footnotetext{\url{https://github.com/epfml/powersgd}}, LSTM\footnote{\url{https://github.com/pytorch/examples/tree/master/word_language_model} } and GNMT\footnote{\url{ https://github.com/mlcommons/training/tree/master/rnn_translator}}.
The hardware configuration and model training details are listed in table~\ref{tab:exp_setting}.

\begin{figure*}[h]
	\centering
	\includegraphics[width=0.5\linewidth]{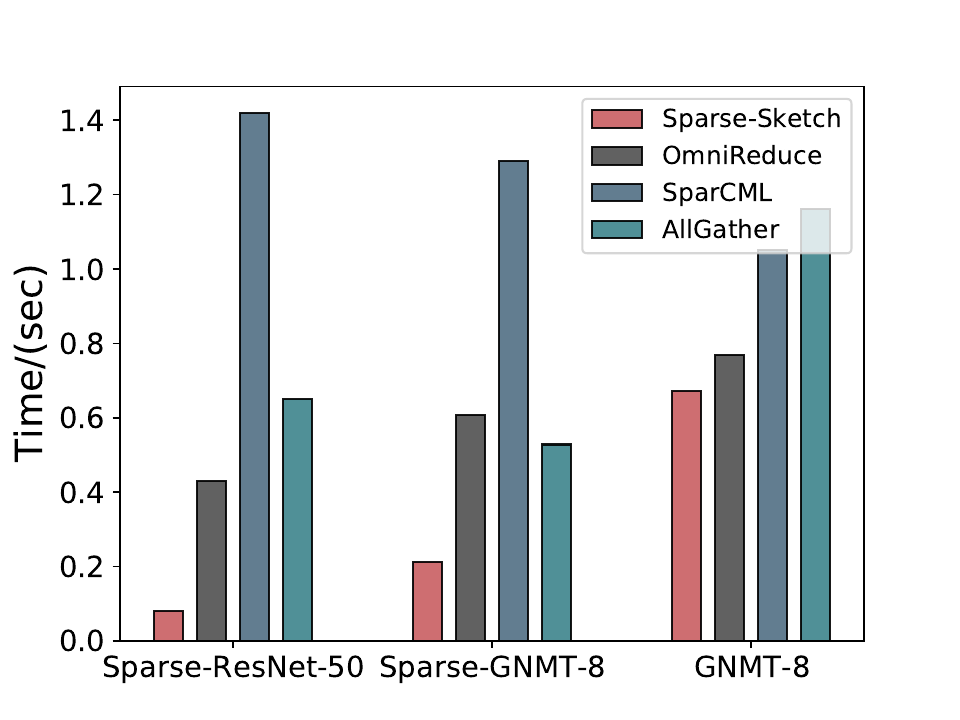}
	\caption{The communication overhead of sparse gradient when training models on 16-GPU cluster. The first two models are ResNet-50 and GNMT-8 with block-wise sparsification, and the last one is GNMT-8, where gradient of embedding layers are sparse.}
	\label{fig:sparsecommtime_appen}
\end{figure*}

In this study, we compare the overhead of different collective communication methods. The results are illustrated in Figure~\ref{fig:sparsecommtime_appen}. Our method reduce 94\% of the communication overhead compared to SparCML, and 87\% to AllGather on ResNet-50 with block-wise sparsification. When applying sparsification on GNMT-8, S2 Reducer decreases the overhead by 83\% and 59\% compared with SparCML and AllGather respectively.

\subsection{Sensitivity Analysis of Hyper-parameter $ \lambda $}{\label{appen:sensitivity}}

\begin{figure}[h]
	\centering
	\begin{subfigure}[b]{0.4\linewidth}
		\includegraphics[width=\linewidth]{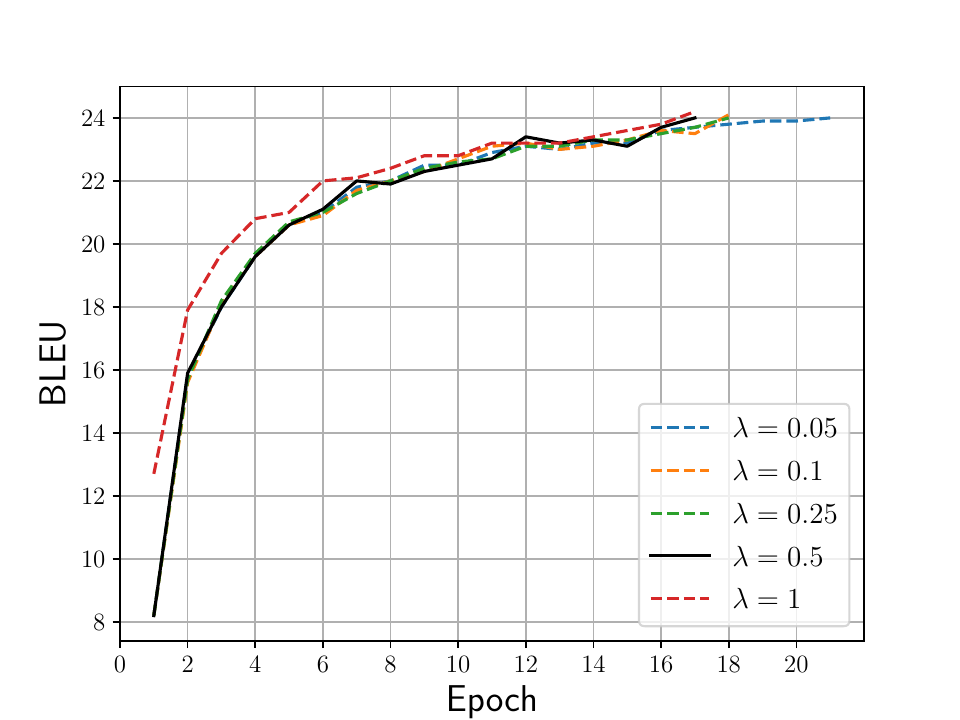}
		\caption{BLEU \textit{Vs.} epoch of GNMT-8}
	\end{subfigure}
	\begin{subfigure}[b]{0.4\linewidth}
		\includegraphics[width=\linewidth]{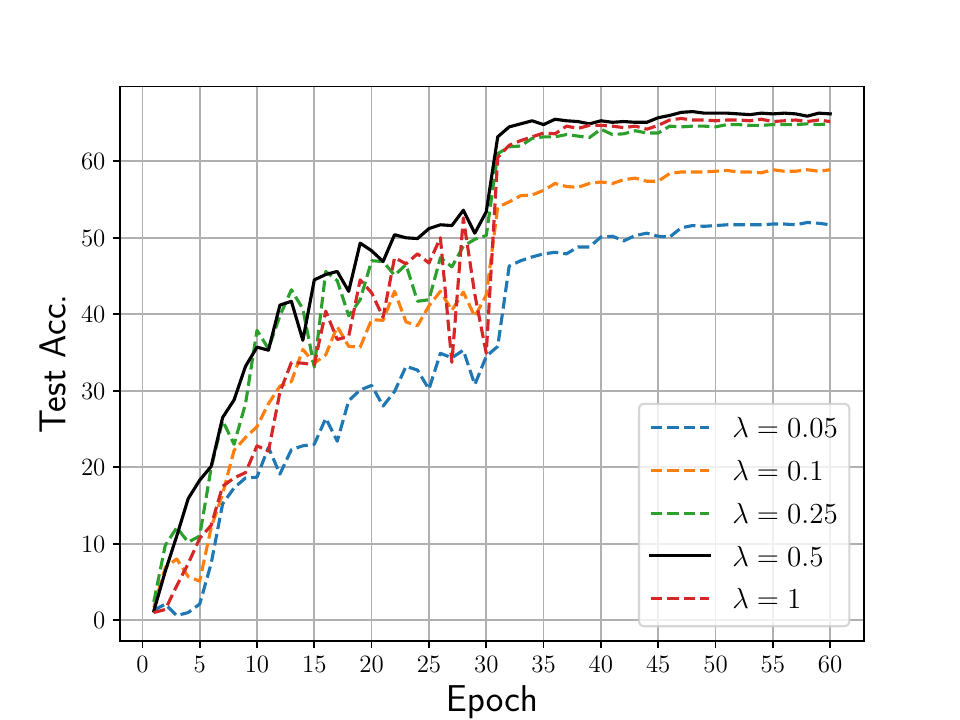}
		\caption{Accuracy \textit{Vs.} epoch of ResNet-50}
	\end{subfigure}
	
	\caption{The model performance of GNMT-8 and block-wise sparsification ResNet-50 when varying $ \lambda $.}
	
	\label{fig:sensitivity}
\end{figure}

Here, we illustrate the model performance when we decrease the size of sparse sketch. The models are trained in a 16-GPU cluster, and the training configuration is the same as Section~\ref{appen:sparse_time} except the hyper-parameter $ \lambda $. As mentioned in Section~\ref{subsec:sparse_sketch}, we represent the ratio of the size of sketch to number of non-zeros with $ \lambda $. In the experiments, $ \lambda $ varies from 1 to 0.05, and both GNMT-8 and ResNet-50 results show that $ \lambda=0.5 $ is the best choice. In fact, the model performance downgrades and the number of epochs to achieve training target increases when we decrease $ \lambda $ from 0.5, while the gains in communication overhead is quiet slight.

\end{document}